**Examining Perceptions Of Astronomy Images Across Mobile Platforms**

Lisa F. Smith[a], Kimberly K. Arcand[b], Jeffrey K. Smith[a],
Randall K. Smith[b], Jay Bookbinder[b], Megan Watzke[b]

[a] University of Otago College of Education, Te Kura Akau Taitoka, 145 Union Street East, PO Box 56, Dunedin, New Zealand
[b] Harvard-Smithsonian Center for Astrophysics, 60 Garden Street, Cambridge, MA 02138, USA

## Abstract

Modern society has led many people to become consumers of data unlike previous generations. How this shift in the way information is communicated and received – including in areas of science – and affects perception and comprehension is still an open question. This study examined one aspect of this digital age: perceptions of astronomical images and their labels, on mobile platforms. Participants were *n* = 2183 respondents to an online survey, and two focus groups (*n* = 12 astrophysicists; *n* = 11 lay public). Online participants were randomly assigned to 1 of 12 images, and compared two label formats. Focus groups compared mobile devices and label formats. Results indicated that the size and quality of the images on the mobile devices affected label comprehension and engagement. The question label format was significantly preferred to the fun fact. Results are discussed in terms of effective science communication using technology.

## Context

Astronomy is a visual science. Even if simply gazing at the night sky with the unaided eye, most people have had some experience with visually processing astronomical information. Many engaging results of modern astronomy come from astronomical sources and phenomena not detectable by the human eye, with astrophysical research expanding into all types of electromagnetic radiation, from radio to infrared, X-ray, and beyond. Astronomical images viewed today are inherently digital, with tens and even hundreds of terabytes of data available from a single observatory archive.[1]

In that respect, astronomy is an ideal subject to be at the forefront of the transformation of how many people in society access and interact with information in today's world. As opposed to generations past or even a decade or so ago, our habits of seeking and consuming data have shifted dramatically. Rather than looking for a physical representation of prior knowledge (e.g., a book in a public library) or news in a relatively updated medium (e.g., the daily newspaper), many people around the world are now accustomed to immediate access to vast amounts of data and imagery on devices such as smartphones and tablets.

Astronomy is also an interesting topic to examine because a large proportion of the undertakings in astronomy are funded by taxpayer dollars. As such, many in the professional astronomical community believe in the responsibility of releasing meaningful results as a form of proof or "public report."[2] The images produced from astronomical data are often a great asset in assuring the public that their money has been well spent in the service of science. They also provide a principal source of information that forms public conceptions about space.[3]

Thus, there is an obligation on the part of those in the professional astronomical community to produce images that are not only meaningful in terms of satisfying public interests, but that accurately communicate the science that resides within the images to greater non-expert audiences.[4]

These visual expressions of the astrophysical research have the potential to be powerful tools of science communication.[5]

Miller[6-9] has established that many adults engage with online sources to obtain scientific information, often related to a sense of wanting to know more. Given that the public can now access a wide range of astronomical images online and across multiple platforms, it becomes even more important to explore what affects the comprehension and perception of those images. A recent Pew Internet Report on mobile usage reported that as of May 2013, 91% of American adults have a cell phone; 56% of American adults have a smartphone; 28% of cell owners own an Android; 25% own an iPhone; 4% own a Blackberry; and, 34% of American adults own a tablet computer.[10] The 2013 Pew Report also noted that 60% of cell phone owners access the Internet, and about 1/3 "of cell internet users say that they mostly use their phone to access the Internet, as opposed to other devices like a desktop, laptop, or tablet computer." Sixty percent of mobile news consumers reported that they routinely visited 2-5 web sites for their news, according to a 2010 Pew Report.[11] The Pew Research Center's Project for Excellence in Journalism noted that 62% of cell phone owners reported using their phone weekly to read up on news.[12] It is reasonable to expect that as the number of mobile devices increases, so will access to those devices as sources of scientific information.

Proposed U.S. federal policies in science, technology, engineering, and math (STEM) education and outreach have called for evidence-based approaches to improve program effectiveness (see e.g., http://www.whitehouse.gov/sites/default/files/microsites/ostp/stem_stratplan_2013.pdf, p. 45). However, no research to date has systematically examined how the public views astronomy images across digital platform sizes. As a starting point, therefore, we have drawn on the scholarly literature on informal science learning, aesthetics, and in expert/novice differences. Research has established that informal learning is often voluntary in nature and self-motivated.[13-15] When people go to museums, they typically attend to the displays and exhibitions that interest them, and stay as long as that interest is maintained.[16] The desire to educate on the part of museum staff needs to be considered alongside the interest levels and backgrounds of the visitors. Thus, a delicate balance is needed[17-18] to explore how people interact with astronomical displays in museums and science centers. In other words, there is a need to understand the attractiveness of the exhibition and the nature of and manner in which the science is communicated. With regard to aesthetics, there are a number of studies that have explored how individuals visually process art[19], how the information accompanying art influences its perception[20], and how experts differ from novices in the perception of art[21]. Finally, expert/novice differences have a long history in cognitive psychology[22] and need to be taken into account when presenting images to the public. In a large online survey, the authors demonstrated that color, explanation, and scale affected the perception and comprehension of astronomical imagery for both experts and novices. They found that expert participants preferred brief, technical explanations for the images as compared to the explanations written with questions or narratives, which proved more effective in engaging those with less scientific background.[23]

Research is needed that establishes how experts and novices differ in their perceptions of images across digital platforms, and what types of information - and in what formats - most enhance the appreciation and understanding of these images on digital platforms. This empirical study addresses these issues, using methods similar to those in the authors' study[23], to systematically examine the relationship between information and user comprehension across digital platforms. The research questions for this study were:

1. How engaging are astronomical images across different types of mobile platforms?
2. What types of information most enhance the comprehension of astronomical images across different mobile platforms?

These questions were explored in an online survey and in two focus groups.

# Method

## Participants

Participants for the online survey were a convenience sample, solicited as volunteers using a mobile device. Participation was sought by posting notices to astronomy related web sites including http://chandra.si.edu and http://apod.nasa.gov/, as well as through social media including Facebook and Twitter accounts for Chandra and the Smithsonian (such as @chandraxray, @sitesExhibits, and @SIAffiliates). In total, there were 2183 useable responses. Of these, 21.4% (*n* = 468) were female and 78.6% (*n* = 1,715) were male. They reported using six types of mobile platforms, with 1.9% (*n* = 41) using a Palm, 4.4% (*n* = 97) using an iPad, 11.1% (*n* = 242) using a Blackberry, 27.0% (*n* = 590) using an iPhone, 29% (*n* = 633) using non-smartphones, and 26.6% (*n* = 580) using a variety of other types of smartphones. Additional demographic information is shown in Table 1. As Table 1 indicates, the participants tended to be over age 35, with a university-level degree. Although 41.2% (*n* = 900) of the participants listed their occupation as "Other," in most cases, these were refinements of the listed categories, for example, science teacher (rather than the more general category, educator) or plumber (rather than trades).

To determine levels of expertise, participants were asked to rate their knowledge of astronomy on a scale of 1 (complete novice) to 10 (expert). The mean was 5.31 (*SD* = 2.30). To further explore the participants' background in astronomy, a series of questions were asked regarding the atmosphere on the Sun, lunar eclipses, Jupiter's great red spot, the rings of Saturn, comets, supernova remnants, what will happen when the Sun "dies," gamma-ray bursts, black holes, and dark matter. Following a technique first used by Smith and Smith[18], for each item, participants used a 0 to 4-point scale to indicate whether they had never heard of it, had heard of it but knew nothing about it, had a vague understanding of it, could understand it when it was discussed, or could talk intelligently about it. Total scores ranged from 7 to 40, with a mean of 31.43 (*SD* = 5.90). For purposes of analysis, these were grouped into top, middle, and lowest thirds of background (see

Table 2). Finally, participants were asked to consider a list of items pertaining to astronomy and to check all that applied. The items were: hobby, online pastime, amateur avocation, profession, something studied in high school, something studied at university. These were summed; scores ranged from 0 to 5 (see Table 2).

Table 1: *Demographic Information*

| | | Online Participants | | Focus Group/Non-Experts | |
|---|---|---|---|---|---|
| Variable | Categories | *n* | % | *n* | % |
| Age Group | <18 | 80 | 3.7 | -- | -- |
| | 18-24 | 144 | 6.6 | 1 | 9.1 |
| | 25-34 | 302 | 13.8 | -- | -- |
| | 35-44 | 394 | 18.0 | 2 | 18.2 |
| | 45-54 | 492 | 22.5 | 3 | 27.3 |
| | 55-64 | 522 | 23.9 | 2 | 18.2 |
| | 65+ | 249 | 11.4 | 3 | 27.3 |
| Highest Level Of Education | Some High School | 126 | 5.8 | -- | -- |
| | High School | 185 | 8.5 | -- | -- |
| | Some University | 442 | 20.2 | 1 | 9.1 |
| | Undergraduate Degree | 684 | 31.3 | 4 | 36.4 |
| | Postgraduate/Masters | 521 | 23.9 | 5 | 45.5 |
| | Doctorate/MD | 225 | 10.4 | 1 | 9.1 |
| Occupation | Astrophysicist | 59 | 2.7 | -- | -- |
| | Educator | 181 | 8.3 | 2 | 18.2 |
| | Executive/Management | 230 | 10.5 | 1 | 9.1 |
| | Homemaker | 28 | 1.3 | -- | -- |
| | Office | 213 | 9.8 | 4 | 36.4 |
| | Retired | 244 | 11.2 | 1 | 9.1 |
| | Trades | 78 | 3.6 | -- | -- |
| | Sales | 59 | 2.7 | -- | -- |
| | Student | 191 | 0.7 | 1 | 9.1 |
| | Other | 900 | 41.2 | 2 | 18.2 |

Participants for the two focus groups were astrophysicists/astronomers (*n* = 12) from the Harvard-Smithsonian Center for Astrophysics (CfA), who formed the “expert” group and volunteers (*n* = 11) from the Boston area, who formed the “non-expert” group. The volunteers from the public were solicited from boston.com and Harvard Gazette’s online notices, as well as some social media advertising on Twitter and Facebook. There were equal numbers of males and females in the astrophysicists/astronomer group; their ages ranged from 34-65, with a mean age of 56 (*SD* =0.55). Using the scales described above, as might be expected, everyone in the expert group gave a self-rating of 10 for expertise, and their mean for the concepts on the 0 to 4-point scale was 30.83 (*SD* = 1.59). The non-expert group (see Tables 1 and 2) comprised *n* = 7 (63.6%) males and *n* = 4 (36.4%) females. Their mean self-rating for knowledge of astronomy was 3.09 (*SD* = 1.70); they were well educated, with a variety of occupations. The two participants who listed "other" occupations were both employed in the arts, one as an artist and one as a filmmaker. For the astronomy concepts, their mean was 22.64 (*SD* = 12.99) and the sum of the background interest items ranged from 0 to 4.

Table 2: *Participants' Levels of Expertise*

| | | Online Participants | | Focus Group/Non-Experts | |
|---|---|---|---|---|---|
| Variable | Categories | *n* | % | *n* | % |
| Self-rating Knowledge of Astronomy (1 = complete novice, 10 = expert) | 1 | 207 | 9.5 | 2 | 18.2 |
| | 2 | 91 | 4.2 | 2 | 18.2 |
| | 3 | 222 | 10.2 | 3 | 27.3 |
| | 4 | 231 | 10.6 | 3 | 27.3 |
| | 5 | 285 | 13.1 | -- | -- |
| | 6 | 359 | 16.4 | -- | -- |
| | 7 | 426 | 19.5 | 1 | 9.1 |
| | 8 | 242 | 11.1 | -- | -- |
| | 9 | 80 | 3.7 | -- | -- |
| | 10 | 40 | 1.8 | -- | -- |
| Knowledge of Concepts Total Score | 0 | 26 | 1.2 | -- | -- |
| | 1 | 516 | 23.6 | 3 | 27.3 |
| | 2 | 900 | 41.2 | 4 | 36.4 |
| | 3 | 741 | 33.9 | 4 | 36.4 |
| Background Items Total Score | 0 | 194 | 8.9 | -- | -- |
| | 1 | 688 | 31.5 | 7 | 63.6 |
| | 2 | 608 | 27.9 | -- | -- |
| | 3 | 458 | 21.0 | 3 | 27.3 |
| | 4 | 186 | 8.5 | 1 | 9.1 |
| | 5 | 49 | 2.2 | -- | -- |

**Materials**

An online survey was developed for this study, comprising a total of 34 items and using 12 astronomical images chosen to represent a full spectrum of object type, scale, and distance from Earth (from images of a planet, comet and our own Sun, to dead and dying stars, black holes, galaxies and clusters of galaxies), as well as electromagnetic radiation type (including data from the visible, infrared, X-ray and ultraviolet portions of the spectrum). The image data came from NASA's Chandra X-ray Observatory, Hubble Space Telescope, Solar Dynamics Observatory, Cassini spacecraft, Spitzer Space Telescope, and GALEX mission, the European Space Observatory and Atacama Pathfinder Experiment (APEX) telescope, and astrophotographers Akira Fujii and Dan Schechter, used with permission. The 12 images used were: Sun, Lunar Eclipse, Saturn, Comet Hale-Bopp, G292.0+1.8, Cat's Eye, Cassiopeia A, Galactic Center, Whirlpool Galaxy, NGC 4696, Bullet Cluster, and Centaurus A. The images themselves were not ones that were in common use at the time of the survey, even though what was depicted (e.g., the Sun, Saturn) would have been familiar. Two types of labels were used for each image, based on the authors' previous findings[23]. One format had a leading question that was answered in the label; the other format was in the form of a narrative with a "fun fact" intended to aid retention of the information. An example of one image and its labels that were used in both the online study and in the focus groups are shown in Figure 1. For the focus groups, a demographics questionnaire and a semi-structured protocol were prepared.

**Procedure**

The study was deemed exempt from ethical review by Harvard University's Institutional Review Board. For the survey, participants were advised that submission of the survey implied consent. They were then randomly assigned to one of the 12 images. Given the length of the online survey and concerns over potential carryover effects, it was decided to have participants respond to only one image. They first responded to a series of both quantitative and open-ended items regarding what they liked or did not like about the image. Next, they were shown one of the two label types with the image, and asked a series of questions designed to explore their comprehension of the label content. They were then presented with the second type of label and asked to compare the two labels in terms of which was more engaging, which best conveyed the information about the image, and which was more likely to "stick" in terms of future retention of information. Next, a series of items asked what the participant still wanted to know about the image, whether he or she was familiar with the image, and to describe how carefully he or she had read the two labels. The survey concluded with the demographic items.

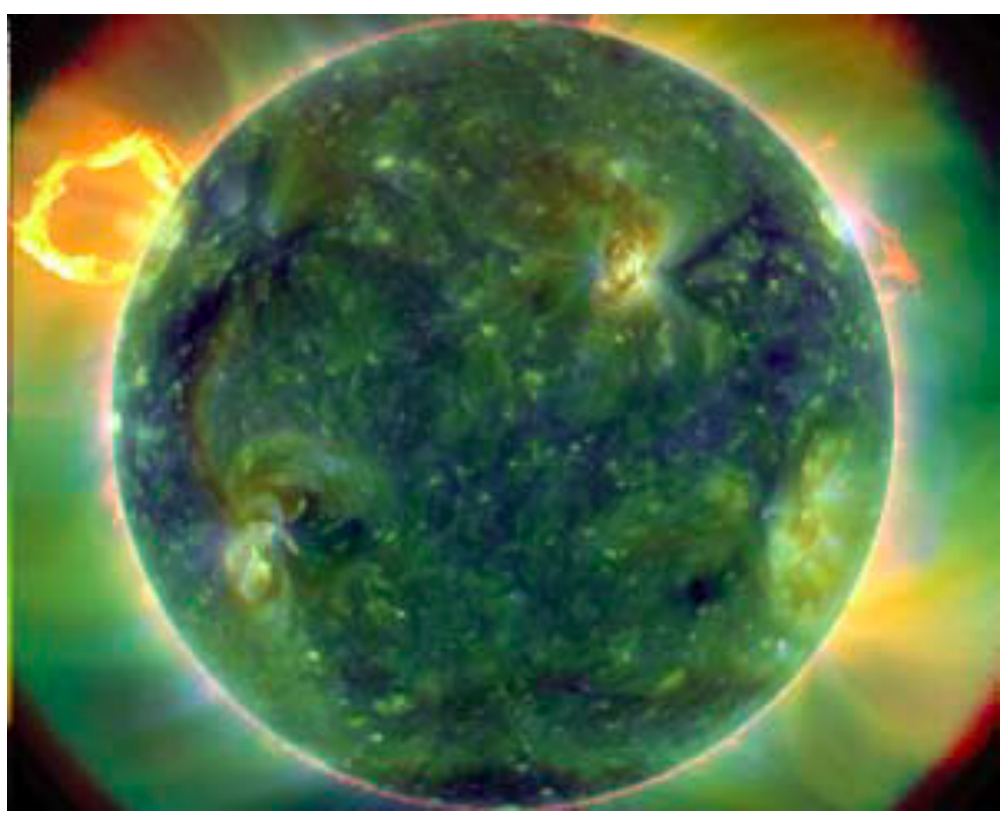

Image of the Sun

*(Question and Answer Format)*
What is a storm on the Sun like? Most of the time when we think of "weather," we are referring to the state of the atmosphere that gives us rain, wind, and changes of temperature. Like the Earth, the Sun has an atmosphere that produces weather. Its weather, however, reaches far beyond the Sun itself into the Solar System. When the Sun is stormy, it results in streams of energized particles that create, among other effects, the Aurora Borealis on Earth. This image, taken in ultraviolet light, shows the turbulent atmosphere of the Sun, where the loops are constantly changing areas of energy and magnetism that drive the Sun's weather.

*(Fun Fact Format)*
Instead of automatically blaming your carrier the next time your cell phone drops a call, maybe think about how the Sun might be responsible. The Sun's surface is constantly changing, generating massive loops of energy and magnetism that are seen in this image taken in ultraviolet light. Occasionally, a giant loop is created that breaks open, sending a stream of energized particles toward the Earth. Sometimes these outbursts are so powerful that they disrupt satellites in orbit, subjecting phone and telephone service to the whims of our nearest star. This image shows the Sun on a typical day, where its turbulent atmosphere goes largely ignored here on Earth.

*Figure 1. Image of the Sun with Question and Answer and Fun Fact labels.*

All focus groups took place following the survey and were audiotaped. Participants were advised that participation implied consent. The intent of the focus groups was to gain a deeper understanding of the survey results through facilitated discussion and hands-on experimentation.

Both expert and non-expert focus groups were presented with images of the Sun, G292, and Centaurus A, across three platforms of different scales and image quality: a high-quality large projection screen, a tablet (iPad), and a smartphone (iPhone). Due to time constraints and concerns regarding repetition, it was not possible to use all 12 images; the three chosen were thought to be representative of the full set of images. The focus group protocol is shown in Appendix A. The groups also responded to the questions as presented in the online survey. They then compared the images using iPads and iPhones. Snacks were provided for both focus groups. In addition, the non-expert participants were given materials from the CfA including postcards, bookmarks, and posters of astronomical images. No other compensation was provided.

**Analysis**

For the online survey, SPSS Version 20 was used to analyze the quantitative data; open-ended items were analyzed using NVivo Version 9. The focus group audiotapes were transcribed by a graduate assistant and compared to notes made by the researchers during the focus groups to collate responses to the research questions.

## Results

The results will be presented by research question, first for the online survey participants and then for the focus groups.

**How Engaging Are Astronomical Images Across Different Types Of Mobile Platforms?**

**Online survey.** A series of items on the survey were used to examine the first research question. Responses to the question, *How much do you like this image?,* was explored in a two-way analysis of variance (ANOVA), with type of mobile platform and image viewed as the independent variables and rating of the image on a scale of 1 (not at all) to 10 (extremely) as the dependent variable. There was no statistically significant interaction, $F$ (55, 2111) = 1.07, *ns* or difference for mobile platform, $F$ (5, 2111) = 1.27, *ns*; however, there was a significant difference for image viewed that yielded a small estimate for an effect size, $F$ (5, 2111) = 2.82, $p < .01$, partial *eta*

squared = .01. A Scheffé post-hoc procedure indicated that NGC4696 was rated significantly lower all other images and the image of Saturn was rated significantly higher than all other images. The means and standard deviations from this analysis are shown on Table 3.

Table 3: *Means and Standard Deviations for Images Viewed in ANOVA for Liking Image By Mobile Platform*

| Image | *M* | *SD* |
|---|---|---|
| NGC 4696 | 5.87 | 2.39 |
| Galactic Center | 6.29 | 2.67 |
| Bullet Cluster | 6.35 | 2.52 |
| Lunar Eclipse | 6.91 | 2.08 |
| G292.0+1.8 | 6.91 | 2.12 |
| Comet Hale-Bopp | 6.93 | 2.16 |
| Sun | 6.93 | 2.30 |
| Whirlpool Galaxy | 7.06 | 2.35 |
| Cassiopeia A | 7.09 | 2.23 |
| Centaurus A | 7.24 | 2.47 |
| Cat's Eye | 7.28 | 2.29 |
| Saturn | 7.51 | 2.19 |

The same independent and dependent variables were analyzed again as an analysis of covariance, using the measures of expertise (self-rating, concepts score, and background score) as covariates, which yielded similar results in that only image viewed was significant. To explore this further, the data were split by image viewed and the measures of expertise were correlated with rating for liking each image. Using alpha < .01, no significant correlations were obtained. Similarly, there were no significant differences for the item, *How familiar were you with this image before you came to this web site?,* although within each image there were some participants who reported having seen that image prior to encountering it in the survey.

To further explore the level of engagement with the images, participants responded to two open-ended items, *What do you like about it?* and *Is there anything you don't like about it?* The responses to these items provided more insight to issue of what worked and what didn't work in viewing the images. Given the lack of statistically significant differences among the platforms, the analyses of these two questions are presented for the total sample.

For the item, *What do you like about it?,* comments centered on seven categories. There were 751 comments about the colors used in the images. Participants were especially positive about the colors used for the image of the Sun, commenting on how it increased their engagement with the image. One participant wrote, “I like the different zones of color. It feeds my interest.” Another stated, “The green color base makes the features easy to see.” Among the comments on color, 21 participants mentioned the use of false color. One comment, about the Sun, was, “Excellent false color imaging!” A total of 415 comments centered on the aesthetic appeal/pleasingness of the images. One participant who viewed the Cat’s Eye commented, “It is so aesthetically pleasing!” Another, responding to Centaurus A, stated, “It’s visually intriguing and aesthetically appealing.” Participants made 320 comments about the detail and composition of the images. For the image of Saturn, one participant noted, “I like the detail of the rings.” Another, viewing Cassiopeia A, stated, “I like the detail, composition, geometry, color” Perceived movement prompted 226 comments, including, “It shows progression, movement” (Lunar Eclipse) and “The movement, it looks like two things colliding" (Bullet Cluster). There were 191 responses that were exclamations of the awesome nature of the images and their overall mysteriousness. In looking at the Whirlpool Galaxy, one participant wrote, “Awesome; the fact that something like this exists!!!!” Another, looking at the Galactic Center, simply said, “Mysterious!” The size of the image was noted by 91 participants. In looking at the Sun, one participant stated, “It’s apparently of massive size, even on a smartphone.” Looking at Saturn, one participant wrote, “I like the fact that it puts in context the size of Saturn’s rings.” The final category included 67 comments on the complexity of the images. One participant responding to G292, stated, “I love the complexity.”

Although some participants responded, “nothing” to the item *Is there anything you don’t like about it?* and 434 participants left that item blank, there were three major categories to the responses. The most frequent response concerned the size of the image, in particular that it was too small. There were 1,265 responses of “too small” across all of the images. When more explanation was given, participants wrote “too small to really appreciate scientific achievement” (Sun) and “too small for iPad – makes it hard to see” (G292). The second category, with 359 comments, addressed the resolution and detail of the images: “At that resolution, the detail is getting smudged away” (Lunar Eclipse) and “low resolution/looks grainy” (Comet Hale-Bopp). The third category related to color, with 156 participants commenting on how the colors appeared on their mobile devices. One stated, “Color of the image. For a small image, a more vibrant photo could probably be found” (Cassiopeia A) and another commented that, “Colors seem ‘washed’” (Galactic Center).

**Focus groups.** The same questions were posed to the focus groups for the three images used. Recall that the expert and non-expert focus group participants viewed three images: the Sun, G292, and Centaurus A, across three platforms: a high-quality large projection screen, iPads, and iPhones. As with the online participants, all participants expressed positive reactions to the images. As would be expected, the experts were familiar with the images and identified colleagues who had worked on creating them, discussed decisions made regarding colors used, and acknowledged their aesthetic appeal. The non-expert group were not familiar with the images presented but all identified the Sun and two recognised Centaurus A. As with the authors’ 2010 finding[23], the non-experts’ sense of wonder was the primary reaction. Comments about the science underlying the images or particulars about their composition followed.

Unlike the online participants, the focus groups had the opportunity to compare the images on two mobile platforms. In both groups, it was clear that size mattered. Viewing the images without any labels, all participants preferred the larger iPad screen to the iPhone. One non-expert summed it up as, “They look good on the iPhone but given a choice, there’s no comparison. I’ll take the iPad!”

**What Types Of Information Most Enhance The Comprehension Of Astronomical Images Across Different Mobile Platforms?**

Using the first label shown, responses to the item, *How much does this label help you to understand the image?* was explored using a two-way analysis of variance (ANOVA), with type of mobile platform and image viewed as the independent variables, and rating of the item a scale of 1 (not at all) to 10 (extremely) as the dependent variable. There were no statistically significant differences for the interaction or either main effect. Using the same type of analysis, there were no significant differences for the items, *How much does this label help you to understand the image?, How well do you think you could explain what this image is showing to another person, having read this label?,* and *How important was this label to your appreciation of this image?*

Significant differences were found, however, when comparing the two labels. Again, the data were split according to which image was viewed. Respondents were asked, *Which one do you think is more interesting/engaging?, Which one do you think conveys the information best about the image?,* and *Which one is more likely to stick with you--so you'll remember the information after you leave this web site?* Using Pearson chi-square statistics, the question and answer label format was compared to the fun fact label format for the images.

For the item, *Which one do you think is more interesting/engaging?,* the question and answer label format (59.2%) was preferred significantly more than the fun fact label (40.8%), at $p < .001$. The fun fact label was chosen only for Saturn and G292. For the item, *Which one do you think conveys the information best about the image?,* the question and answer label format (66.1%) again was preferred significantly more than the fun fact label (33.9%), at $p < .001$. The fun fact label format was chosen only for Saturn. We can only speculate on the reason for this. It may be because the fun fact, which was, "The rings of Saturn are one of astronomy's most famous objects, first observed by Galileo Galilei with his telescope more than 400 years ago," resonated due to its personal link with Galileo and the beginning of the use of telecopes.

For the item, *Which one is more likely to stick with you--so you'll remember the information after you leave this web site?*, the question and answer label format (62.1%) was preferred significantly more than the fun fact label (37.9%), at $p$ <.001. The fun fact label format was chosen for Saturn and the Bullet Cluster.

Respondents were asked, *In general, how carefully did you read the labels that were with the image?* Only 13.7% of the sample stated that they had skimmed the labels; 48.1% read them carefully, and 38.3% read them very carefully. A reanalysis of the data using responses from only those who read the labels carefully or very carefully did not alter the results.

**Focus groups.** The same items were asked of the non-expert focus group participants. The expert focus group participants were not asked to compare the labels, as the aim of this part of the study was to examine the label formats for the general public. The non-experts preferred the question and answer label format over the fun fact label format for all three of the images that they viewed. One participant noted, “I can’t resist wanting to know the answer to the question.” Another said, “The fun facts are interesting but they are almost a bit too patronizing, like I’m being talked down to.”

When the images were presented paired with the labels on the iPhones and iPads, the clear preference was for the iPad. Comments included, “You can hardly read the label on the iPhone.” “Even if I increase the type size on the iPhone, then I have to scroll along to read and that’s a strain.” “It’s still pretty small type on the iPad, but it’s a lot easier to read than on the iPhone.”

**Questions From the Online Participants**

A final item asked of the online participants was, *What do you still want to know about this image?* A total of 1,636 participants asked one or more questions. Across the images, the questions primarily concerned six topics. By far, the most frequently occurring question asked how the image was captured ($n$ = 532). This was typically asked in a straightforward manner, for example, “How was the image generated?” “What telescope or satellite captured the image?” The second asked for more specific scientific details, for example about the duration of the flares for the Sun, or the composition of the rings of Saturn (*n* = 484). Examples are, “What is the size/scope of the flare or CME?” “What is the duration of the phenomenon seen [on the Sun]?” and “How do the composition of the rings relate to Saturn’s moons?” The third category related to the colors in the image (*n* = 216), for example, “What are the real colors of the rings?” and “are the colors as they would appear to a local human eye?” The fourth category involved questions about the dimensions of the image and a desire for more information about the scale of the image in comparison to the Earth (*n* = 173). These included, “What are the dimensions of supernova, remnant stars, etc.?” and “What is the basic dimension and scale to compare with known measurements?” The fifth category comprised questions about exact location of the image, distance from other astronomical bodies, and in the case of the Hale-Bopp Comet, the Sun, and the Lunar Eclipse, how often what was depicted could be seen (*n* = 134). Examples are, “What is its location in relation to other space objects?” “How far is it from Earth?” and “How often might we see something like this [Hale-Bopp Comet]?” Finally, the sixth category was made up of requests for information regarding where to purchase a copy of the image, how to obtain a high-resolution version of the image, and how to download it (*n* = 45), for example, “How can I download a high-resolution copy of this image?” and “Can I purchase this image?”

## Conclusions

This study set out to examine how different types of mobile platforms affect viewing astronomical images, and to explore what types of information most enhance the comprehension of astronomical images across different mobile platforms. The key finding in terms of mobile platforms was that when participants were permitted to engage in comparisons across mobile platforms in the focus groups, the size of the mobile platform mattered. Bigger was better. In the online survey, where participants viewed one image on a single mobile platform, the mean ratings for a Likert item about liking the image ranged from roughly 6 to 8 on a scale of 1 (low) to 10 (high), indicating a good level of satisfaction. However, comments in an open-ended item indicated the size and resolution of the image, particularly on mobile phones, were problematic. This finding provides support for Smith and Smith's concept of facsimile accommodation in that, as might be expected, in the absence of a comparison, participants tended to adapt to the mobile platform size.[24] There were few expert/non-expert differences in the findings reported here; more research is needed in this area.

When considering what types of information most enhances comprehension of astronomical images across different mobile platforms, the findings from this study indicate that labels beginning with a question that is then answered are preferred to labels that are written in a less formal, more conversational manner, making use of a "fun fact." This finding supports previous work by the authors on these types of label formats for astronomical images[23], as well as Hohenstein and Uyen Tran's study[25], where questions in the labels in a science museum setting generated more conversation among visitors to two of three exhibitions examined.

Differences between experts and novices were not as pronounced as they had been in previous work[23]; however, the focus in this study was more on comparing platforms and not looking at differences in responses. In this study, both experts and novices reported that the larger platform was greatly superior to the smaller one.

There are, as always, limitations to the findings reported here. Generalisation beyond these types of images and formats for labels is not recommended. The sample sizes for the public focus groups were small and were comprised of a convenience sample from one geographic area rich in universities. It should be noted, however, that this study recently has been replicated *in situ* across four major museums, with results forthcoming.

The results from this study suggest that both the size and the quality of the images on mobile platforms, as well as the accompanying text, need to be considered in communicating the science that underpins deep space imagery. Bringing the skies to individuals' mobile devices is not the issue. The issue is how to do this such that the public is engaged, informed, and eager for more. We live in a world in which technological changes are rapid and far-reaching. This study provides some insight as to how we can make the most of exciting technology in the service of science communication and education.

## Acknowledgements

The authors gratefully acknowledge support for this project from a Smithsonian Institution Scholarly Studies Program Grant, # 40488100HH0003, with additional support from the Hinode X-ray Telescope, performed under NASA contract NNM07AB07C, and the Education and Outreach group for NASA's Chandra X-ray Observatory, operated by the Smithsonian Astrophysical Observatory (SAO) under NASA Contract NAS8-03060. The authors also thank Jerry Bonnell and Robert J. Nemiroff, the authors of the NASA Astronomy Picture of the Day web site.

**Appendix A**

## Protocol for Focus Groups

*For each question, prompt participants to expand on their responses and not simply provide ratings.*

For Each Image, Ask:

1. What you think of this image? Do you like it? Use a scale of 1 (not at all) to 10 (extremely).

2. What do you like/is there anything you don't like about it?

Show The First Label

3. How much does this label help you to understand the image? Use a scale of 1 (not at all) to 10 (extremely).

4. How well do you think you could explain what this image is showing to another person, having read this label? Use a scale of 1 (not at all) to 10 (completely)

5. How important was this label to your appreciation of the image? Use a scale of 1 (not at all) to 10 (extremely)

Show The Second Label

6. Which label do you think is more interesting/engaging? The first label or the second label?

7. Which one do you think conveys the information best about the image? The first label or the second label?

8. Which one is more likely to stick with you – so you'll remember the information after you leave the exhibition? The first label or the second label?

9. What do you still want to know about this image?

**Authors' Biographies**

**Lisa F. Smith** is Professor of Education and Dean of the University of Otago College of Education. Her research has a focus on the psychology of aesthetics. Contact: University of Otago College of Education, 145 Union Street East, Dunedin 9054, New Zealand; <lisa.smith@otago.ac.nz>, +64 3 479 9014

**Kimberly K. Arcand** is Visualization Lead for NASA's Chandra X-ray Observatory, Chandra X-ray Center/Harvard-Smithsonian Center for Astrophysics, 60 Garden Street, Cambridge, MA, USA <kkowal@cfa.harvard.edu> +1 617 218 7196

**Jeffrey K. Smith** is Professor of Education and Associate Dean-Research of the University of Otago College of Education. Contact: University of Otago College of Education, 145 Union Street East, Dunedin 9054, New Zealand; <jeffrey.smith@otago.ac.nz>, +64 3 479 5467

**Randall K. Smith** is an Astrophysicist at the Harvard-Smithsonian Center for Astrophysics. Harvard-Smithsonian Center for Astrophysics, 60 Garden Street, Cambridge, MA, USA <rsmith@cfa.harvard.edu>

**Jay Bookbinder** is a Senior Astrophysicist and Program Manager at the Harvard-Smithsonian Center for Astrophysics. Harvard-Smithsonian Center for Astrophysics, 60 Garden Street, Cambridge, MA, USA <jbookbinder@cfa.harvard.edu>

**Megan Watzke** is a Media & Communications Specialist at the Chandra X-ray Center/Harvard-Smithsonian Center for Astrophysics. Harvard-Smithsonian Center for Astrophysics, 60 Garden Street, Cambridge, MA, USA<mwatzke@cfa.harvard.edu>